\documentclass[aps,superscriptaddress,eqsecnum,nofootinbib,showpacs,preprintnumbers]{revtex4-2}
\usepackage{amsmath}
\usepackage{amssymb}
\usepackage{amsfonts,amsthm,bm}
\usepackage{color,xcolor}
\usepackage[greek,english]{babel}
\usepackage{titlesec}

\usepackage{graphicx,epsfig}
\graphicspath{{figures/}{./}}
\usepackage{float}
\usepackage{subfigure}
\usepackage{tikz}
\newcommand{\m}{\mu}

\newcommand{\be}{\begin{eqnarray}}
	\newcommand{\ee}{\end{eqnarray}}
\newcommand{\bea}{\begin{eqnarray}}
	
	\newcommand{\eea}{\end{eqnarray}}

\def\m{\mu}

\def\Ref{\ref}

\definecolor{azure(colorwheel)}{rgb}{0.0, 0.5, 1.0}
\definecolor{DarkViolet}{RGB}{148,0,211}
\definecolor{MyDarkBlue}{rgb}{0,0.1,0.7}
\definecolor{DarkBlue}{RGB}{0,0,153}
\definecolor{amber}{rgb}{1.0, 0.49, 0.0}
\definecolor{amaranth}{rgb}{0.9, 0.17, 0.31}
\definecolor{nicered}{rgb}{0.7,0.1,0.1}
\definecolor{brown}{rgb}{0.5,0.1,0.1}
\definecolor{nicegreen}{rgb}{0.0,0.3,0.0}
\definecolor{tealgreen}{rgb}{0.0, 0.51, 0.5}

\definecolor{tclr}{RGB}{148,0,211}

\usepackage{hyperref}
\hypersetup{colorlinks,bookmarksopen,
	bookmarksnumbered,
	citecolor={nicered},
	linkcolor={MyDarkBlue},
	urlcolor={tealgreen},
	pdfstartview=FitH}

\newcommand{\beq}{\begin{equation}}
\newcommand{\eeq}{\end{equation}}
\newcommand{\bseq}{\begin{subequations}}
	\newcommand{\eseq}{\end{subequations}}

\usepackage{orcidlink}
\def\idbako{\orcidlink{0000-0002-3012-6144}}
\def\idchri{\orcidlink{0000-0001-5852-514X}}
\def\idpapa{\orcidlink{0000-0003-1244-922X}}
\def\idchat{\orcidlink{0000-0003-4479-2970}}

\begin{document}

\title{Stealth Ellis Wormholes in Horndeski Theories}

\author{Athanasios Bakopoulos\idbako}
\email{a.bakop@uoi.gr}
\affiliation{Physics Department, School of Applied Mathematical and Physical Sciences,
	National Technical University of Athens, 15780 Zografou Campus,
	Athens, Greece.}
 \affiliation{Division of Applied Analysis, Department of Mathematics,
University of Patras, Rio Patras GR-26504, Greece.}

\author{Nikos Chatzifotis\idchat }
\email{chatzifotisn@gmail.com}
\affiliation{Physics Department, School of Applied Mathematical and Physical Sciences,
	National Technical University of Athens, 15780 Zografou Campus,
	Athens, Greece.}

\author{Cristian Erices\idchri}
\email{cristian.erices@ucentral.cl}
\affiliation{Vicerrectoría Académica,  Universidad  Central de Chile, Toesca 1783, Santiago 8320000, Chile}
\affiliation{Departamento de Matemática, Física y Estadística, Universidad Católica del Maule, Av. San Miguel 3605, Talca 3480094, Chile}

\author{Eleftherios Papantonopoulos\idpapa }
\email{lpapa@central.ntua.gr}
\affiliation{Physics Department, School of Applied Mathematical and Physical Sciences,
	National Technical University of Athens, 15780 Zografou Campus,
	Athens, Greece.}

\begin{abstract}
	\noindent In this work we are revisiting the well studied Ellis wormhole solution in a Horndeski theory motivated from the Kaluza-Klein compactification procedure of the more fundamental higher dimensional Lovelock gravity. We show that the Ellis wormhole is analytically supported by a gravitational theory with a non-trivial coupling to the Gauss-Bonnet term and we expand upon this notion by introducing higher derivative contributions of the scalar field. The extension of the gravitational theory does not yield any back-reacting component on the spacetime metric, which establishes the Ellis wormhole as a stealth solution in the generalized framework. We propose two simple mechanisms that dress the wormhole with an effective ADM mass. The first procedure is related to a conformal transformation of the metric which maps the theory to another Horndeski subclass, while the second one is inspired by the spontaneous scalarization effect on black holes. 
\end{abstract}

\maketitle

\flushbottom

\tableofcontents

\section{Introduction}
Wormholes are one of the simplest and most exotic static solutions of Einstein's equations. The throat of a wormhole is able to connect two space-times, or sometimes two distant parts of the same universe. Conceived initially as a hypothetical structure, and after Flamm provided its mathematical notion \cite{Flamm}, wormholes were rediscovered by Einstein and Rosen as a bridge-like structure known as the Einstein-Rosen bridge \cite{ER}. After Wheeler first coined the term `wormhole' \cite{Wheeler}, Ellis proposed a geodesically complete wormhole or `drainhole' by introducing a scalar field minimally coupled to gravity \cite{Ellis:1973yv}. However, it was not until the seminal article from Morris and Thorne that the first humanly traversable wormhole was mathematically derived \cite{MorrisThorne}. From asymptotically flat \cite{Visser:1989kg,Visser:1989kh,Visser:1995cc} to asymptotically (anti)de-Sitter wormholes \cite{Lemos:2003jb}, a major drawback is inherent in these solutions in the theory of general relativity (GR). The so-called ﬂare-out condition, which is the condition that confers and maintains the geometrical structure of the wormhole, is non-compatible with the Null Energy Condition (NEC), violating all the other conditions \cite{Visser:1995cc}. It turned out that a scalar field of phantom nature, i.e. with an opposite sign in front of its kinetic term, was sufficiently exotic to support traversable wormholes \cite{Ellis:1973yv,Bronnikov:1973fh}. This long quest for physically meaningful wormhole solutions has found a prolific theoretical framework in modified theories of gravity.

A number of wormhole solutions have been derived in modified theories of gravity. A remarkable result was found in \cite{Kanti:2011jz}, where the authors construct a traversable wormhole without needing any form of exotic matter.
This is in fact a consequence of the introduction of quadratic terms in the curvatures at the level of the action, which allows to satisfy the flare-out conditions without spoiling the NEC. In this context, exotic matter is reduced or even not needed when wormholes solutions are realized on alternative
theoretical frameworks. Traversable wormholes have been found in many modified gravitational theories, such as in $f(R)$ gravity \cite{Lobo:2009ip,Mishra:2021xfl,Mazharimousavi:2016npo,Sokoliuk:2022xcf,Solanki:2023onp,Kavya:2023lms,Karakasis:2021tqx}, $f(R,T)$ gravity \cite{Banerjee:2020uyi,Sahoo:2019aqz,Sahoo:2019ffc,Sahoo:2020sva}, modified teleparallel gravity \cite{Boehmer:2012uyw}, non-minimal couplings \cite{MontelongoGarcia:2010xd,Garcia:2010xb}, extra fundamental fields \cite{Harko:2013yb, DeFalco:2020afv,DeFalco:2021btn,DeFalco:2021klh,DeFalco:2021ksd,DiGrezia:2017daq}, Einstein-Gauss-Bonnet gravity \cite{Bhawal:1992sz,Dotti:2007az,Mehdizadeh:2015jra}, Einstein-Scalar-Gauss-Bonnet gravity \cite{Antoniou:2019awm,Bakopoulos:2020mmy}, Brans-Dicke theory \cite{Agnese:1995kd,Anchordoqui:1996jh,Papantonopoulos:2019ugr}, Randal-Sundrum model \cite{MaldacenaWH}, braneworld configurations \cite{Lobo:2007qi}, metric-Palatini gravity \cite{Capozziello:2012hr,Rosa:2021yym}, with thin shells \cite{Lobo:2020vqh,Berry:2020tky,Mehdizadeh:2015dta,PhysRevD.101.124035, Chatzifotis:2022mob}, Einsteinian Cubic gravity \cite{Mehdizadeh:2019qvc}, Einstein-Dirac-Maxwell \cite{Kain:2023ann}, massive gravity \cite{Dutta:2023wfg} and by disformal transformations \cite{Chatzifotis:2021hpg,  Bakopoulos:2021liw, Bakopoulos:2022gdv, Babichev:2022awg, Bakopoulos:2022csr}, just to name a few during the last decade. For further details, the reader may refer to a comprehensive review in \cite{Alcubierre:2017pqm}.

During the 70's Horndeski built the most general scalar-tensor theory with equations of motion of second order for both the metric and for the scalar field \cite{Horndeski}, therefore avoiding Ostrogradsky ghosts. Although Horndeski theory was forgotten for decades, great interest resurfaced in the community when it was proved that the decoupling limit of the DGP model \cite{DGP} corresponds to the Galileon theory \cite{Galileon}. This  theory is symmetric under galilean transformation ---hence its name---  of the scalar field $\phi\rightarrow\phi+c+b_{\m} x^{\m}$ ($c$, $b_{\m}$ being constant) and although it is constructed in flat space, possesses novel properties. A remarkable one is its Vainshtein mechanism having useful applications in cosmology since it allows the effects of the scalar field to be significant at cosmological scales but not at smaller ones \cite{Vainshtein}. Then, the covariantization of this theory was achieved \cite{Deffayet1,Deffayet2} leaving behind the shift symmetry, and getting the so called generalized Galileon theory which was proven to be equivalent to the Horndeski theory \cite{Kobayashi}. Moreover, in four dimensions, Horndeski theory can be obtained through the compactification process of the higher dimensional Lovelock theory \cite{Charmousis:2012dw}. The generality of Horndeski theory provide a number of sectors with remarkable properties. For a complete review refer to \cite{Kobayashi:2019hrl}.

As expected, the majority of analytical solutions for compact objects in Horndeski theories have been obtained within the subclass of theories that enjoy parity symmetry $\phi\rightarrow-\phi$ and/or shift symmetry $\phi\rightarrow\phi+\text{cst}$ \cite{Cisterna:2014nua,Erices:2015xua,Bakopoulos:2022gdv, Babichev:2017guv,Babichev:2016fbg, Babichev:2016rlq, Charmousis:2015aya,Brihaye:2016lin,Corral:2021tww}. However, under physical grounds there are no compelling reasons to restrict ourselves to theories with such symmetries, and in consequence, this freedom opens the possibility to explore  spacetime geometries with quite different properties than GR \cite{Charmousis:2014zaa,Antoniou:2017hxj,Bakopoulos:2019tvc,Bakopoulos:2020dfg,Bakopoulos:2021dry,Bakopoulos:2023hkh,Chatzifotis:2022ubq,Chatzifotis:2022ene,Cisterna:2018hzf,Erices:2021uyu,Karakasis:2023ljt,Karakasis:2023hni,Karakasis:2022fep,Karakasis:2021rpn,Erices:2022bws,Anabalon:2012tu,Barrientos:2022avi,Cisterna:2023uqf}. In this scenario, there is a potential risk of dealing with much more involved field equations, precluding the existence of analytical solutions. It has been found that such a theory \cite{Babichev:2023dhs} is derived from the generalization of the potentials appearing in Kaluza-Klein compactifications originating from Lovelock theory \cite{Charmousis:2012dw}. This particular sector of Horndeski theories can be identified as effective string theory actions with an IR logarithmic running
for the dilaton, higher order corrections and generalised Liouville type potentials.

Motivated by a lack of such symmetries, in this work, we proceed to study wormhole solutions in a Horndeski theory framework. Our first approach is to consider  a phantom  scalar field which is non-minimally coupled to the Gauss-Bonnet (GB) term. Solving the field equations of this  theory and demanding that the redshift function is constant, we find that the Ellis wormhole is a stealth solution of the theory. The charge of the scalar field determines the throat of the Ellis wormhole, as expected. It is well known that in four dimensions the GB term is a topological term. In our case, despite the fact that the scalar field is non-trivially coupled to the GB term, there is no contribution of the GB term to the background metric. Indeed, by calculating the corresponding topological current of the GB term in a completely general spherically symmetric metric ansatz, we show that if the redshift function is constant, the GB term vanishes identically. It is in this sense that the Ellis wormhole can be supported as a stealth solution of the theory.

Following up on this, we extend our gravitational action by introducing higher derivative terms, such as a non-trivial derivative coupling to the Einstein tensor and higher order derivative couplings of the scalar field. These terms can be generated in the Horndeski theory from the generalization of the potentials in the Kaluza-Klein compactification, originating from higher dimensional Lovelock gravity, as previously mentioned. We found that despite the presence of non-trivial derivative couplings, the Ellis wormhole is still supported by the extended theory, thus appearing as an effective stealth solution.

We then turn our attention on whether the stealth Ellis wormhole in the presence of higher curvature terms and higher derivatives of the scalar field can be extended beyond its stealth nature. Firstly, we show that performing a conformal transformation, which maps us to a different Horndeski subclass, a massive wormhole solution can be generated in the conformal frame. Thus, the Ellis solution, originally a non-gravitating object, acquires an effective ADM mass. On the other hand, constraining ourselves onto the original gravitational theory, we propose a new method of obtaining massive wormhole solutions.  In particular, since an actual Birkhoff theorem is absent in the framework of the Horndeski theories, one may construct novel solutions in a perturbative manner by allowing the higher order terms to back react on the stealth solution. This procedure resembles the spontaneous scalarization effect on black holes \cite{Doneva:2018rou, Guo:2020sdu}. However, since the stealth wormholes already carry a non trivial scalar charge we distinguish ourselves from this phenomenon by coining the term ``rescalarization''. 

The paper is organized as follows: In Sec. \ref{sec2} we extract the analytic stealth wormhole solution for Scalar-Gauss-Bonnet gravity. Then, in Sec. \ref{sec3}, it is shown that this configuration is admitted as a stealth solution of the more generic theory obtained as the generalization of the potentials in the Kaluza-Klein compactification of higher dimensional Lovelock gravity. Additionally, the massive configuration is derived through a suitable conformal transformation and some geometric properties are described. Sec. \ref{sec4} is devoted to construct the massive wormhole of the theory by solving the field equations perturbatively around the stealth wormhole. We review our results and conclude in Sec. \ref{sec5}.

\section{Stealth Ellis Wormhole in Gauss-Bonnet gravity}
\label{sec2}

It is well known that the Ellis wormhole solution described by the line element
\begin{equation}
	\label{ellis}
	ds^2=-dt^2+dx^2+(x^2+x_0^2)d\Omega^2,
\end{equation}
where $x_0$ denotes the throat radius of the wormhole, is a solution to the simple extension of the Einstein-Hilbert action by the addition of a purely phantom scalar field \textbf{with the configuration of $\Phi(x)=\text{arctan}(x/x_0)$}. The fact that the corresponding redshift function of the geometry is vanishing, which implies a zero mass for the wormhole, enables one to re-extract the solution in the context of a more general scalar-tensor theory. Our first approach shall be to consider the extension of the gravitational theory with a Gauss-Bonnet coupling term, i.e.,
\begin{equation}
	\label{theory1}
	S=\int d^4x \frac{\sqrt{|g|}}{16\pi G}\left[R-2 \delta (\partial\Phi)^2+V[\Phi]\mathcal{G}\right]\ ,
\end{equation}
where $\mathcal{G}=R_{\alpha\beta\mu\nu}R^{\alpha\beta\mu\nu}-4R_{\mu\nu}R^{\mu\nu}+R^2$ is the Gauss-Bonnet term. The $\delta$ component of the gravitational theory may assume values of $\pm1$ in order to differentiate between a phantom and canonical nature of the matter content, while $V[\Phi]$ is a coupling function of the scalar field to the Gauss-Bonnet term to be determined by the assumption that the Ellis wormhole is a solution to the theory. The corresponding equations of motion by the variation of the theory with respect to the metric and the matter content are
\begin{align}
	\label{eom}
	G_{\mu\nu}=T_{\mu\nu}+\Theta_{\mu\nu}\\
	\Box \Phi=-\frac{\delta}{4} \dot V (\Phi)\mathcal{G}\ ,
\end{align}
where 
\begin{equation}
	\label{stresskinetic}
	T_{\mu\nu}=2\delta \left(\nabla_\mu\Phi\nabla_\nu\Phi-\frac{1}{2}g_{\mu\nu}(\partial\Phi)^2\right)\ ,
\end{equation}
is the stress-energy tensor associated with the kinetic term of the matter content, and,
\begin{equation}
	\label{stressgb}
	\Theta_{\mu\nu}=-2\nabla_\alpha\left[P^{\alpha\,\,\,\beta}_{\,\,\,\mu\,\,\, \nu}\nabla_\beta V(\Phi)+(\mu\leftrightarrow\nu)\right],
\end{equation}
with $P_{\alpha\beta\gamma\delta}=-\frac{1}{4}\epsilon_{\alpha\beta\mu\nu}R^{\mu\nu\kappa\lambda}\epsilon_{\kappa\lambda\gamma\delta}$ denoting the double dual of the Riemann tensor, is the stress-energy tensor associated with the variation of the Gauss-Bonnet extension and $\dot V (\Phi)$ denotes the derivative of $V(\Phi)$ with respect to $\Phi$. 

The addition of the Gauss-Bonnet term was not accidental. Indeed, it is well known that in four dimensions, the Gauss-Bonnet term is purely topological and, as such, can be expressed in terms of the divergence of the topological current

\begin{equation}
	\label{top}
	K^\xi=-\varepsilon^{\xi\alpha\beta\gamma}\varepsilon^{\,\,\,\rho\,\,\,\nu}_{\sigma\,\,\,\mu\,\,\,}\left[\Gamma^{\sigma}_{\,\,\,\alpha\rho}\partial_\beta\Gamma^\mu_{\,\,\,\gamma\nu}+\frac{2}{3}\Gamma^\sigma_{\,\,\,\alpha\rho}\Gamma^\mu_{\,\,\,\beta\lambda}\Gamma^\lambda_{\,\,\,\gamma\nu}\right]
\end{equation}
Assuming a most general spherical symmetric ansatz of the form
\begin{equation}
	\label{ansatz}
	ds^2=-F(x)dt^2+\frac{dx^2}{F(x)}+r^2(x)d\Omega^2,
\end{equation}
one may immediately verify that the topological current of (\Ref{top}) reads
\begin{equation}
	\label{topresult}
	K^\xi=-\left\{0,\frac{2 F'(x) \left(1-2 F(x) r'(x)^2\right)}{r(x)^2},-\frac{2 \cot (\theta ) F''(x)}{r(x)^2},0\right\}\ ,
\end{equation}
where prime denotes the derivative with respect to $x$. Since $\nabla_\xi K^\xi=\mathcal{G}$, constraining ourselves to a vanishing redshift function, i.e. $F(x)=1$, the Gauss-Bonnet term will always vanish. Therefore, fixing the $F(x)$ degree of freedom on the local solution space of the theory, 
\begin{equation}
	\label{reducedmetric}
	ds^2=-dt^2+dx^2+r^2(x)d\Omega^2\ ,
\end{equation} 
thus remaining with 3 degrees of freedom ($r(x),\Phi(x),V(\Phi)$), the equations of motion described in (\Ref{eom}) are effectively reduced to
\begin{align}
		\label{redcuedeom}
	G_{\mu\nu}&=T_{\mu\nu}+\Theta_{\mu\nu}\ ,\\
	\Box \Phi&=0\ .
\end{align}
The scalar field degree of freedom can be immediately solved under our reduced metric ansatz. Indeed, from the $\theta\theta$ component of the gravitational equations of motion, one easily obtains that
\begin{equation}
	\label{eom33}
	r''(x)+\delta  r(x) \Phi '(x)^2=0\ .
\end{equation}
The above equation states that there are 2 possible solutions depending on the sign of $\delta$, i.e. whether the scalar field is canonical or phantom. For the case of $\delta=+1$, the above equation contains positive definite components and the sole (acceptable) solution is the trivial Minkowski spacetime with a vanishing scalar field. On the other hand, setting $\delta=-1$, which will be our focus for the remaining of our work, one immediately obtains that
\begin{equation}
	\label{scalar}
	\Phi(x)=\Phi_0\pm \int \sqrt{\frac{r''(x)}{r(x)}}dx\ .
\end{equation}
The constant term can always be absorbed by a simple field redefinition. Consequently, one may always choose the solution of
\begin{equation}
	\label{scalarsol}
	\Phi(x)=\int \sqrt{\frac{r''(x)}{r(x)}}dx\ .
\end{equation} 
Having obtained the scalar solution, the corresponding $xx$ component of the gravitational equations of motion is a simple differential equation of $r(x)$,
\begin{equation}
	\label{throat}
	r(x) r''(x)+r'(x)^2-1=0\rightarrow r(x)=\sqrt{(x-c_1)^2+x_0^2}\ ,
\end{equation}
where $c_1$ and $x_0$ are integration constants. It should be stressed at this point that the $c_1$ ``charge'' is a simple translation in the radial coordinate and can always be absorbed with no physical implications. Indeed, setting $\ell^2=(x-c_1)^2$, one obtains $d\ell^2=dx^2$. Therefore, the throat radius function of the wormhole reads
\begin{equation}
	\label{throatsol}
	r(x)=\sqrt{x^2+x_0^2}\ ,
\end{equation}
which yields that the scalar field is indeed of the Ellis wormhole form
\begin{equation}
	\label{Phisol}
	\Phi(x)=\arctan\left(\frac{x}{x_0}\right)\ ,
\end{equation}
which naturally satisfies $\Box\Phi=0$. Finally, we deal with the $tt$ gravitational equation of motion, which can be inverted to a differential equation of $V(\Phi)$ with respect to $\Phi$ and reads
\begin{equation}
	\label{coupling}
	\ddot V (\Phi)-4 \tan (\Phi) \dot V (\Phi)=0\rightarrow V(\Phi)=\alpha  \left(\frac{2 \tan (\Phi)}{3}+\frac{1}{3} \tan (\Phi) \sec ^2(\Phi)\right)\ .
\end{equation}
This concludes the first section of our work, where we have shown that the Ellis wormhole solution described by the doublet
\begin{equation}
	\label{doublet}
	\left\{g_{\mu\nu}=diag (-1,1,(x^2+x_0^2),(x^2+x_0^2)\sin^2\theta),\Phi(x)=\arctan\left(\frac{x}{x_0}\right)\right\}\ ,
\end{equation} is indeed supported by the extended gravitational theory
\begin{equation}
	\label{theory}
	S=\int d^4x\frac{\sqrt{|g|}}{16\pi G}\left[R+2(\partial\Phi^2)+\alpha\left(\frac{2 \tan (\Phi)}{3}+\frac{1}{3} \tan (\Phi) \sec ^2(\Phi)\right)\mathcal{G}\right]\ ,
\end{equation}
with a nontrivial coupling to the Gauss-Bonnet term.
\section{Wormholes in Gauss-Bonnet gravity with Higher order derivative couplings}\label{sec3}

\subsection{The Stealth Ellis configuration}

Following our previous result, our goal in this section is to generalize our process with the addition of higher order derivative couplings. It is well known that the presence of a phantom scalar field in a gravitational action will generate instabilities on the local solutions supported by the theory \cite{Bronnikov:2012ch}. However, when the action contains higher order contributions on the scalar field kinetic term, one cannot immediately assume that the $\delta=-1$ configuration will yield unstable solutions, since it is probable that the higher order kinetic term contributions are able to rectify the instability pathologies associated with a phantom scalar field \cite{Chatzifotis:2021pak}. In particular, due to the higher order derivative couplings, it is unclear from the definition of the conjugate momentum of the scalar field whether the matter content is indeed of phantom nature.  To this end, we are considering the following further extension of the gravitational theory
\begin{equation}
	\begin{split}
	\label{extendedtheory}
	S=\int d^4x\frac{\sqrt{|g|}}{16\pi G}\left[R+2(\partial\Phi^2)+\alpha\left(\frac{2 \tan (\Phi)}{3}+\frac{1}{3} \tan (\Phi) \sec ^2(\Phi)\right)\mathcal{G}\right.\\
	\left.	+V_2(\Phi)G^{\mu\nu}\partial_\mu\Phi\partial_\nu\Phi+V_3(\Phi)(\nabla\Phi)^4+V_4(\Phi)(\nabla\Phi)^2(\Box\Phi)\right],
\end{split}
\end{equation}

\noindent which was motivated by the work done in \cite{Babichev:2023dhs}, where they considered a generic Horndeski gravity occuring from the generalization of the potentials in the Kaluza-Klein compactification originating from higher dimensional Lovelock gravity. Naturally, the complexity of the above theory leads to highly non-trivial equations of motion, which hold no real merit in being written down explicitly. Our goal is to test whether the Ellis wormhole doublet described in (\Ref{doublet}) can be supported by the theory in a similar manner to the work done in the previous section. 

It turns out that the simplicity of the Ellis scalar field solution, allows the entire gravitational equations of motion system to be inverted and re-expressed as differential equations of the 3 potentials with respect to the scalar field. Indeed, the three gravitational equations of motion in the orthonormal frame are found to read as follows,
\begin{align}
    \label{tt}
    tt:&\quad 2 \tan (\Phi) \dot{V}_2(\Phi)-V_2(\Phi) \left(6 \tan ^2(\Phi)-3\right)-V_3(\Phi)+\dot{V}_4(\Phi)-4 V_4(\Phi) \tan (\Phi)=0\ ,\\
    \label{xx}
    xx:&\quad V_2(\Phi) \left(2 \tan ^2(\Phi)-1\right)+3 V_3(\Phi)-\dot{V}_4(\Phi)+4 V_4(\Phi) \tan (\Phi)=0\ ,\\
    \label{thetatheta}
    \theta\theta:&\quad \tan (\Phi) \dot{V}_2(\Phi)+V_2(\Phi)-4 V_2(\Phi) \tan ^2(\Phi)-V_3(\Phi)+\dot{V}_4(\Phi)-4 V_4(\Phi) \tan (\Phi)=0\ .
\end{align}

It is important to note here that $x_0$ does not appear in the gravitational equations of motion, which implies that the primary charge nature of the throat radius survives in the generalized Horndeski framework with higher order derivative couplings. It is easy to notice that the 
$tt-\theta\theta$ combination of the equations of motion yields a differential equation solely in terms of $V_2$. Indeed, one may easily extract that,
\begin{equation}
    \label{V2}
    tt-\theta\theta:\quad \sin (2 \Phi) \dot{V}_2(\Phi)+4 V_2(\Phi) \cos (2 \Phi)=0\implies V_2(\Phi)=\beta \csc ^2(2 \Phi)\ ,
\end{equation}
\noindent where $\beta$ is an integration constant. Moving on, we also notice that both the $tt$ and the $xx$ equations are linear algebraic equations of $V_3$. From the $xx$ component we find that,
\begin{equation}
    \label{V3}
    V_3(\Phi)=\frac{1}{3} \left(\beta \csc ^2(2 \Phi)-\frac{1}{2} \beta \sec ^4(\Phi)+\dot{V}_4(\Phi)-4 V_4(\Phi) \tan (\Phi)\right)\ .
\end{equation}
Plugging the above result into the $tt$ equation of motion we extract a simple differential equation of $V_4$
\begin{equation}
    \label{V4}
    \beta-2 \sin ^2(\Phi) \cos ^4(\Phi) \dot{V}_4(\Phi)+V_4(\Phi) \sin ^3(2 \Phi)=0\ ,
\end{equation}
whose solution reads
\begin{equation}
    \label{V_4}
   V_4(\Phi)=-\frac{1}{2} \beta \csc (\Phi) \sec ^3(\Phi)+\gamma \sec ^4(\Phi)\ ,
\end{equation}
while consequently, $V_3$ is found to be
\begin{equation}
    \label{V_3}
    V_3(\Phi)=\beta \csc^2(2\Phi)\ .
\end{equation}

We note at this point that this choice of the corresponding potentials yields that the entire extension of the action under consideration vanishes on shell and it is in this sense that the Ellis wormhole is supported as a local solution of the theory. However, any divergence of the trivial Ellis wormhole on the geometry, either by considering massive wormholes or the effect of gravitational waves will switch on the Gauss-Bonnet and the higher derivative terms, which allows for a much richer structure than the initially considered gravitational theory. 

 A second important comment is that, while all wormhole solutions are known to be unstable in the Horndeski framework\cite{Rubakov:2016zah,Evseev:2017jek}, the addition of higher order derivative terms will in principle strongly affect the lifetime of the corresponding solution. A careful and non-trivial stability examination needs to be implemented to verify whether the higher order terms extend the lifetime of the wormhole. Such an analysis is out of scope of this paper. 

\subsection{Massive wormholes via conformal transformations} 

So far, we have been focusing strictly on the trivial massless Ellis wormhole. It should be noted that a corresponding massive configuration can always be found in the general framework of Horndeski gravity by a simple conformal transformation. Indeed, the simplicity of the scalar field solution will be a very helpful aid. We are constraining ourselves to a target spacetime with a $Z_2$ symmetry on the radial coordinate in order to avoid known causal pathologies associated with time-machine configurations, such as the Ellis drainhole \cite{Ellis:1973yv}, and Schwarzschild-like asymptotic behaviour on the $g_{tt}$ component of the metric in order to obtain a positive definite ADM mass at infinity. A way to approach this is to consider the conformal transformation of 
\begin{equation}
    \label{conformal}
    g_{\mu\nu}\rightarrow (1-2\zeta\cos\Phi)g_{\mu\nu}
\end{equation}
where $\zeta$ is a novel parameter that will be constrained in order to have well defined non-degenerate conformal transformation.  The metric we obtain yields the following line-element

\begin{equation}
    \label{ds2}
    ds^2=-\left(1-\frac{2\zeta x_0}{\sqrt{x^2+x_0^2}}\right)dt^2+\left(1-\frac{2\zeta x_0}{\sqrt{x^2+x_0^2}}\right)\left[dx^2+(x^2+x_0^2)d\Omega^2\right].
\end{equation}
In order to extract any meaningful results on the nature of the above spacetime, we should stress at this point that the corresponding area radius function
\begin{equation}
    \label{arearadius}
    R(x)=\sqrt{\left(1-\frac{2\zeta x_0}{\sqrt{x^2+x_0^2}}\right)(x^2+x_0^2)}=\sqrt{x_0 \left(x_0-2 \zeta \sqrt{x^2+x_0^2}\right)+x^2}
\end{equation}
yields that $R(x)\sim x$ at $x\rightarrow\infty$, which implies that we can test the asymptotic behaviour in terms of the $x$ radial coordinate. On the other hand, since we are requesting the absence of horizons, which incidentally yields the conformal transformation to be well-defined, we are constraining the positive definite $\zeta$ parameter to 
\begin{equation}
    \label{zeta}
    \zeta<\frac{1}{2}\ .
\end{equation}
In addition, it is easily tested that the corresponding throat of the wormhole lies at $x=0$, since the area radius function contains a single global minimum there under the constrain of (\Ref{zeta}) with a transformed throat radius  
\begin{equation}
    \label{throat2}
    x_{t}=\sqrt{1-2\zeta}x_0\ .
\end{equation}

It is now trivial to deduce that the corresponding ADM mass of the above wormhole is simply $M_{effective}=\zeta x_0$, which implies that although the scalar field contains a single charge, i.e. the throat radius, it is always possible to construct massive wormholes with a fixed ratio of mass to throat radius defined by the coupling constant we introduced in the conformal transformation. Finally, we would like to comment that since the conformal transformation does not contain any real valued singular points, the absence of singularities in our new spacetime is immediately verified.

\section{Rescalarization of the Ellis wormhole: Massive wormholes}\label{sec4}

The Ellis wormhole stands as a fascinating solution within the general Horndeski theory, offering insights into exotic spacetime structures. However, as a stealth solution, the Ellis wormhole does not provide explicit information regarding the coupling constants that govern the underlying Horndeski theory.  %
This situation closely resembles the scenario wherein a local solution within the framework of General Relativity is underpinned by a scalar-tensor theory, where the scalar field is constrained to be non-dynamical. In this context, a nuanced correlation with the concept of``scalarization" emerges.
Scalarization refers to the emergence of scalar field configurations in the presence of a trivial background solution, significantly impacting the system's properties.
Likewise, the Ellis wormhole can be regarded as a solution within the minimal extension of the Einstein-Hilbert action, featuring a non-trivial phantom scalar field. We presented a generalized extension of the aforementioned theory, ensuring the continued validity of the Ellis solution. This is achieved by virtue of the fact that both the action extension and the corresponding stress-energy tensors  vanish on-shell. 

 Much like the case of natural scalarization\footnote{We should stress that the natural scalarization process is intrinsically different from spontaneous scalarization \cite{Antoniou:2017acq,Doneva:2017bvd,Silva:2017uqg} where the emergence of a non-trivial scalar field background is related to a tachyonic instability.} on black holes, where one lets the switched off terms of the action to back-react on the solution and extract a hairy black hole in a perturbative manner, we wish to follow a similar approach. In particular, by letting the extension of the action to affect the solution,  the existence of a second non-trivial solution alongside the Ellis wormhole becomes a possibility, potentially unveiling crucial details about the Horndeski theory itself.   %
In the case of the Ellis wormhole, which already possesses a distinctive scalar field characterizing its background solution, a modified term, ``rescalarization,'' is introduced to distinguish this scenario from conventional scalarization. Rescalarization reflects the idea that the existing scalar field of the Ellis wormhole may  acquire additional characteristics due to the presence of higher orders in the action, thereby shedding light on the Horndeski theory's fundamental aspects. 

This section delves into the intriguing phenomenon of rescalarization associated with the Ellis wormhole and explores the emergence of massive wormhole solutions within the Horndeski theory framework. By investigating the interplay between the non-trivial scalar field and the second solution, we aim to unravel the unique features and implications of rescalarized, massive wormholes.  
The rescalarization of the Ellis wormhole unveils new avenues for studying exotic spacetime geometries and enriches our knowledge of the diverse gravitational phenomena that can exist in the universe.

In our pursuit of discovering new solutions within the generalized Horndeski theory, as discussed in the previous section, we direct our attention to the specific coupling functions described by Eqs. (\ref{coupling}), (\ref{V2}), (\ref{V_4}), and (\ref{V_3}) in the action given by Eq. (\ref{extendedtheory}). To facilitate our analysis, we adopt the following metric ansatz,
\begin{equation}
\label{ansatz1}
ds^2=-F(x)dt^2+\frac{dx^2}{S(x)}+(x^2+x_0^2)d\Omega^2\ .
\end{equation}
Due to the complexity of the field equations, rendering analytical solutions unfeasible, we focus on the weak field approximation. Within this approximation, we assume that the strength of the theory, characterized by the coupling constant, is small. Thus, we seek perturbative solutions by expanding around a known background solution. Since the Ellis wormhole serves as a recognized stealth solution within the theory, we select it as our background. Consequently, we choose to expand the solutions in terms of the Gauss-Bonnet coupling constant, denoted as $\alpha$. To maintain simplicity and expand solely in one parameter, we assume that the remaining two coupling constants of the theory, $\beta$ and $\gamma$, are proportional to $\alpha$. Specifically, we set $\beta=\alpha\lambda_1$ and $\gamma=\alpha\lambda_2$, where $\lambda_{1,2}$ represent two dimensionless constants.
In terms of the metric functions, we propose the following expansions:
\begin{align}
    F(x)&=1+\alpha f(x) + \mathcal{O}(\alpha^2)\ ,\\[1mm]
    S(x)&=1+\alpha s(x) + \mathcal{O}(\alpha^2)\ ,\\[1mm]
    \Phi(x)&=\arctan\left( \frac{x}{x_0}\right)+ \alpha h(x) + \mathcal{O}(\alpha^2)\ .
\end{align}
These expansions allow us to perturbatively investigate the effects of the higher-order theory coupling constant and its associated scalar field on the metric and scalar functions, providing insights into the existence of new solutions within the Horndeski theory.

Upon substituting the expansions into the field equations and considering the limit as $\alpha\rightarrow 0$, we can derive the first-order equations in $\alpha$,
\begin{align}
& (-2 x_0 h+x s)' =0\ , \\[1mm]
& x f'+2 x_0 h'+s=0\ , \\[1mm]
&(x^2+x_0^2)f'' +x (f'-s')-2 s=0\ .
\end{align}
The above equations correspond to the $(t,t)$ equation, $(r,r)$ equation, and the combination $(t,t)-(\theta,\theta)$, respectively. Remarkably, we notice that the constants $\lambda_1$ and $\lambda_2$ do not appear in the aforementioned equations. Therefore, at first order in $\alpha$, the solutions depend solely on the Gauss-Bonnet term. Integrating the above system yields,
\begin{align}
h(x)&=\frac{h_1 x_0 }{x} \arctan\left(\frac{x}{x_0}\right) -h_1+\frac{h_0}{x}-\frac{s_0}{2 x_0}\ ,\\[1mm]
f(x)&=f_0+2 h_1 \arctan\left(\frac{x}{x_0}\right)\ ,\\[2mm]
s(x)&=\frac{2 x_0}{x^2}\left( - h_1 x+h_1 x_0 \arctan\left(\frac{x}{x_0}\right)+h_0 \right)\ ,
\end{align}
where $h_{0}$, $h_1$, $f_0$, and $s_0$ are integration constants.
To ensure finiteness of the scalar field at $x=0$, we set $h_0=0$, and for the scalar field to exhibit the same asymptotics as the Ellis wormhole at infinity, we set $s_0=-2h_1x_0$. Additionally, the requirement of asymptotic flatness implies $f_0=-\pi h_1$. As a result, there is only one independent integration constant ($h_1$). Thus, the complete solution is given by,
\begin{align}
F(x)&=1+\alpha \left[2 h_1 \arctan\left(\frac{x}{x_0}\right)+\pi h_1\right]\ ,\\[1mm]
S(x)&=1- \frac{2\alpha h_1 x_0}{x^2}\left[ x-x_0\arctan\left(\frac{x}{x_0}\right) \right]\ ,\\[1mm]
\Phi(x)&=\arctan\left( \frac{x}{x_0}\right)+\frac{\alpha h_1 x_0}{x}\arctan\left(\frac{x}{x_0}\right)\ .
\end{align}
This solution is regular at $x=0$ and lacks horizons, thus describing a traversable wormhole. It generalizes the Ellis wormhole within the framework of the generalized Horndeski theory.
By expanding the solution at infinity, we find,
\begin{equation}
F(x)=1-\frac{2\alpha h_1 x_0}{x}+ \mathcal{O}\left(\frac{1}{x^2}\right)\ , \qquad \text{and} \qquad S(x)=1-\frac{2\alpha h_1 x_0}{x}+ \mathcal{O}\left(\frac{1}{x^2}\right)\ .
\end{equation}
Therefore, through the process of rescalarization, the Ellis wormhole, originally a non-gravitating object, acquires an ADM mass term of $M=\alpha h_1 x_0$. We observe that the ADM mass depends on the coupling constant $\alpha$, suggesting that the Gauss-Bonnet term introduces a mass term to the Ellis wormhole.
Finally, from the expansion of the scalar field at infinity, we can deduce the new scalar charge of the solution as $q=-x_0(1- \frac{1}{2}\alpha h_1 \pi) =-x_0 +\frac{1}{2}M \pi.$ This reveals that the Gauss-Bonnet term adds a small contribution to the scalar charge of the new solution, which is associated with the mass of the wormhole. By using the above equation, and expanding in $\alpha$, we may express the novel throat radius $x_0$ in terms of the solution parameters $q,\, h_1$ and $\alpha$ as $|x_0|=q(1+\frac{1}{2}\alpha h_1 \pi).$ Therefore, for  the same value of the scalar charge $q$, the rescalarized wormhole is larger than the original Ellis one.

It should be stressed at this point that, in contrast to the conformal transformation mechanism of extracting massive wormholes, the spacetime configuration that we reach does not seem to have a $Z_2$ symmetry on the radial coordinate. Constraining ourselves to proper Minkowski asymptotics on both universes, we find that the temporal coordinate needs to be differently rescaled in each patch of spacetime. This yields causal pathologies in the solution which can be immediately remedied by taking the absolute value of the $x$ radial coordinate in the solution. Naturally, this implies a non-differentiability issue at the throat, which can be solved by introducing a simple thin-shell of matter. 

\section{Conclusions}\label{sec5}

This research endeavor delves into a novel exploration of wormholes within the context of generalized Horndeski theories, which encompass the Gauss-Bonnet term and additional higher-order terms inspired by the renowned Lovelock theory. Our primary objective was to prove the existence of the Ellis wormhole as a stealth solution in the extended framework and meticulously investigate the rescalarization process on the stealth geometry. Thus, we unveiled the intricate result of an effective ADM mass arising solely as an effect of the higher order terms. 

In the initial section of this study, we focused our attention on the Einstein-scalar-Gauss-Bonnet theory, wherein a phantom scalar field was introduced as an essential component. Through the imposition of a vanishing redshift function and the utilization of a specific metric ansatz, we succeeded in simplifying the intricate equations of motion to a more manageable form. This reduction in complexity provided us with a valuable opportunity to ascertain the scalar field and metric solutions associated with the theory. Our investigation yielded a remarkable outcome: we discovered that, under a particular coupling function, the Einstein-scalar-Gauss-Bonnet theory admits the existence of the Ellis wormhole as a stealth solution. This peculiar result marks a significant milestone as, to our knowledge, it represents the very first instance of obtaining an exact analytic spherically symmetric solution within the framework of the Einstein-scalar-Gauss-Bonnet theory.

In the subsequent section, higher-order derivative couplings are included within the gravitational theory. Thereby, inspired by the compactifications of the Lovelock theory, the theoretical framework is enriched by the introduction of supplementary potentials into the action. Subsequently, we derived the equations of motion governing these potentials. By considering the esteemed Ellis wormhole solution as a background, we successfully reformulated the equations of motion in terms of derivatives of the potentials with respect to the scalar field $\Phi$. Remarkably, we proved the existence of a unique form of the new potentials, which renders the generalized theory capable of accommodating the familiar Ellis wormhole as a stealth solution. 

The rescalarization of the Ellis wormhole within the generalized Horndeski theory has provided intriguing insights into the nature of wormholes and the underlying gravitational theory. By considering the existence of a second non-trivial solution alongside the Ellis wormhole, we have uncovered the versatility and richness of the Horndeski theory, which lacks the limitations imposed by uniqueness theorems in GR. The concept of rescalarization, coined in this study, has allowed us to explore massive wormholes and their implications within the theory.

Through our analysis, we have demonstrated that the Gauss-Bonnet term plays a pivotal role in the rescalarization process. By expanding the solutions in terms of the Gauss-Bonnet coupling constant, we have derived first-order equations that depend solely on this term, revealing its significant influence on the dynamics of the rescalarized Ellis wormhole. This finding highlights the importance of considering higher-order terms in gravitational theories beyond GR to fully comprehend the behavior of exotic spacetime structures.

The rescalarized solution of the Ellis wormhole within the framework of the generalized Horndeski theory manifests compelling characteristics. Specifically, it represents a regular traversable wormhole, which possesses notable properties worthy of investigation. Notably, the inclusion of the Gauss-Bonnet term in the theory leads to the emergence of an ADM mass term within the wormhole structure. This acquired mass term signifies an intriguing connection between the Gauss-Bonnet term and the mass attribute associated with the wormhole. Additionally, an analysis of the scalar charge reveals a modest contribution originating from the Gauss-Bonnet term, linked to the mass of the wormhole.

In summary, we found that the connection of exotic compact objects to the gravitational theories is rather subtle, since the Ellis wormhole may appear as a stealth solution in expanded theoretical frameworks. Moreover, through the process of rescalarization, the investigation of massive wormholes and their properties provides valuable insights into the dynamics and implications of higher-order terms in the theory. This study serves as a stepping stone for future explorations into the theoretical frameworks that support wormholes and possibly other intriguing configurations.

\section*{Acknowledgments}

\noindent The research project was supported by the Hellenic Foundation for Research and Innovation (H.F.R.I.) under the “3rd Call for H.F.R.I. Research Projects to support Post-Doctoral Researchers” (Project Number: 7212). AB  in particular thanks the Physics Department at the Silesian University of Opava for hospitality and support within the project CZ.02.2.69/0.0/0.0/18\_054/0014696. The work of N.C. and E.P. is supported by the research project
of the National Technical University of Athens (NTUA) 65232600-ACT-MTG: {\it Alleviating Cosmological Tensions Through Modified Theories of Gravity}. C.E. is funded by Agencia Nacional de Investigación y Desarrollo (ANID) through Proyecto Fondecyt Iniciación folio 11221063, Etapa 2023. We are very happy to thank 
Theodoros Nakas for encouraging and useful discussion.

\bibliography{Bibliography}{}
\bibliographystyle{utphys}

\end{document}